\documentclass[%
reprint,
superscriptaddress,
showpacs,
amsmath,amssymb,
aps,
prl,
]{revtex4-1}

\usepackage{graphicx}
\usepackage{dcolumn}
\usepackage{bm}

\begin{document}
\DeclareGraphicsExtensions{.pdf,.png,.jpg,.eps,.tiff}

\title{Blueprint for a microwave trapped-ion quantum computer}
\author{B. Lekitsch}
\affiliation{Department of Physics and Astronomy, University of Sussex, Brighton, BN1 9QH, UK}
\author{S. Weidt}
\affiliation{Department of Physics and Astronomy, University of Sussex, Brighton, BN1 9QH, UK}
\author{A. G. Fowler}
\affiliation{Google INC, Santa Barbara, CA 93117, USA}
\author{K. M\o lmer}
\affiliation{Department of Physics and Astronomy, Aarhus University, DK-8000 Aarhus C, Denmark}
\author{S. J. Devitt}
\affiliation{Center for Emergent Matter Science, RIKEN, Wakoshi, Saitama 315-0198, Japan}
\author{Ch. Wunderlich}
\affiliation{Department Physik, Naturwissenschaftlich-Technische Fakult\"at, Universit\"at Siegen, 57068 Siegen, Germany}
\author{W. K. Hensinger}
\affiliation{Department of Physics and Astronomy, University of Sussex, Brighton, BN1 9QH, UK}
\email{W.K.Hensinger@sussex.ac.uk}

\begin{abstract}
The availability of a universal quantum computer will have fundamental impact on a vast number of research fields and society as a whole. An increasingly large scientific and industrial community is working towards the realization of such a device. An arbitrarily large quantum computer is best constructed using a modular approach. We present a blueprint for a trapped-ion based scalable quantum computer module which makes it possible to create a scalable quantum computer architecture based on long-wavelength radiation quantum gates. The modules control all operations as stand-alone units, are constructed using silicon microfabrication techniques and they are within reach of current technology. To perform the required quantum computations, the modules make use of long-wavelength-radiation based quantum gate technology. To scale this microwave quantum computer architecture to an arbitrary size we present a fully scalable design that makes use of ion transport between different modules, thereby allowing arbitrarily many modules to be connected to construct a large-scale device. A high-error-threshold surface error correction code can be implemented in the proposed architecture to execute fault-tolerant operations. With only minor adjustments the proposed modules are also suitable for alternative trapped-ion quantum computer architectures, such as schemes using photonic interconnects.
\end{abstract}

\maketitle

\section{Introduction}

Trapped atomic ions are a very promising candidate for the realization of a universal quantum computer having demonstrated robust, high-fidelity state preparation \cite{Blatt1, Harty1, Wunderlich1} and readout \cite{Harty1, Noek1}, high-fidelity universal gate operations \cite{Monroe1, Akerman1, Harty1} and long qubit coherence times \cite{Harty1, Timoney1}.

The concept of using ion transport on microfabricated trap arrays to realize an ion trap quantum computer was proposed by Kielpinski et al. \cite{Kielpinski1}. Developing a comprehensive quantum computing architecture with trapped ions has since attracted a lot of interest. Recently, an approach addressing this was developed by Monroe et al. \cite{Monroe2, Monroe3}, where ion trap modules, also called elementary logic units (ELUs), are photonically interconnected using commercial fibres and optical crossconnect switches. This proposal demonstrates that going from one to many modules is within reach of current technology and thereby provides an interesting path to a large-scale ion trap quantum computer. Theoretical investigations of this approach have shown that it can be used for fault-tolerant quantum computing even in the presence of noisy and lossy links \cite{Monroe3} when combined with entanglement purification \cite{Nickerson1}.

An important challenge towards building a large-scale trapped-ion quantum computer still remains: the development of a detailed blueprint for the individual modules which need to be capable of performing all required fundamental quantum operations and ideally act as a stand-alone small-scale quantum processor. Each module must also offer efficient connections with additional modules to create a universal quantum computer architecture.

In previously proposed trapped-ion quantum computing architectures, modules are powered by laser-driven single and multi-qubit gates. However, the vast amount of individually controlled and stabilized laser beams required in such an architecture would make the required engineering to build a large-scale quantum computer challenging. Here we propose an architecture which is based on a concept involving global long-wavelength radiation and locally applied magnetic fields \cite{Seb1}. The gate interactions are based on a mechanism first proposed by Mintert and Wunderlich in 2001 \cite{Mintert1} making use of magnetic field gradients within dedicated gate zones. Only global laser light sheets are required for loading, Doppler cooling and state preparation and readout of ions, while laser driven quantum gates requiring careful alignment in each gate zone are not required in our approach. Large scale quantum computers, that rely on laser gates and are capable of solving classically intractable problems may require millions of individual laser beams that have to be precisely aligned with respect to individual entanglement regions and need to be individually controlled. In our microwave-based architecture all laser fields except the readout beams are supplied as sheets and do not have to be precisely aligned or individually controlled.

We present the blueprint for a scalable microwave trapped-ion quantum computer module, which is based on today's silicon semiconductor and ion trap technology. The modules, driven by global laser and microwave fields, perform ion loading, ion shuttling, generate locally addressable magnetic fields as well as magnetic field gradients to perform single and multi-qubit gates and feature on-chip photo detectors for state readout. All gate, shuttling and state readout operations are controlled by on-chip electronics and a cooling system is integrated into the module to allow for efficient temperature management. Each module, when placed in an ultra-high vacuum system and powered by global laser and microwave fields operates as a modular stand-alone quantum computer.

Architectures based on photonic interconnects have great potential for scaling up quantum computing \cite{Monroe2, Monroe3}, however the interaction rate between modules and therefore the speed of a computer based on these is typically slow \cite{Hucul1} compared to the execution time of other quantum operations \cite{Noek1, Ballance1, Harty1}. We propose an alternative method of scaling to a large number of modules based on technology which aligns modules next to each other, enabling ion transport between adjacent modules. A universal two-dimensional architecture is formed by fast transport \cite{Bowler1, Walther1} of qubits from one module to adjacent modules. A suitable high-error-threshold error correction code that only relies on nearest-neighbour interactions was developed by Fowler et al. \cite{Fowler1} and can be implemented using this architecture.

The addition of photonic interconnect regions constitutes only a minor adjustment to this blueprint, therefore our microwave trapped-ion quantum computer modules can also be connected using photonic interconnects making the modules useful for alternative architectures proposed so far \cite{Monroe2, Monroe3}.

Our manuscript is structured as follows. A brief overview of microwave-based quantum operations is given in Sec. II. The blueprint for individual quantum computer modules is provided in Sec. III. In Sec. IV we discuss how modules can be connected and used to create a universal quantum computer of arbitrary size. In Sec. V the implementation of the surface error correction code within this architecture is discussed and the execution of a 2048 and 1024 bit number Shor factoring algorithm is described, identifying appropriate resources and resulting specifications.

\section{Microwave-based quantum gates}

Single- and multi-qubit gates, executed with high-fidelity, are essential building blocks of a universal quantum computer. For trapped-ion quantum computing, internal states of atomic ions serve as qubits, while the Coulomb interaction between closely spaced ions makes conditional quantum gates with two or more qubits possible \cite{Blatt1}. In order to couple the dynamics of internal qubit states and motional states, and thus implement multi-qubit gate operations, precisely aligned laser beams have predominantly been used. This has lead to the experimental demonstrations of multi-qubit gates with up to 14 ions \cite{Monz1} and the demonstration of a two-qubit gate in the fault-tolerant regime \cite{Ballance1}. Nevertheless, there are challenges with the above implementations when trying to scale them up to a large number of qubits and when trying to increase the gate fidelity further to reduce the overall system size. Technical challenges when operating the large number of laser beams required for a large-scale quantum computer system include intensity- and phase fluctuations, frequency drifts of the laser output, micron-precise beam alignment, beam pointing instabilities and non-perfect beam quality. In addition to the technical challenges, off-resonant coupling to states outside of the qubit subspace when using Raman beams can pose an additional challenge.

A promising solution to the stability and scalability challenges that come with using lasers to implement large-scale multi-qubit gate operations was proposed by Mintert and Wunderlich in 2001 \cite{Mintert1} and makes use of microwave radiation in conjunction with a static magnetic field gradient. Microwave radiation has since been used to perform single qubit gates with unprecedented fidelity \cite{Harty1, Brown1}, featuring an error per gate as low as $10^{-6}$ \cite{Harty1} and when combined with locally adjustable magnetic fields or magnetic field gradients, individual addressing of closely spaced ions has been demonstrated with crosstalk as low as $10^{-5}$ \cite{Piltz1}.

Coupling between internal states of trapped ions and their motion, necessary for multi-qubit gates, is induced by electromagnetic radiation. Due to the long wavelength (on the order of centimetres) this coupling is vanishingly small for free-running microwaves and is, thus, not useful on its own for multi-qubit gate operations. However, when adding a static magnetic field gradient, which exerts a force due to the magnetic moment associated with the qubit states of the trapped ion, multi-qubit gates can indeed be implemented \cite{Mintert1}. Such magnetic field gradient induced coupling was first used to implement a two-qubit gate between nearest and non-nearest neighbours by Khromova et al. \cite{Khromova1}.

Besides using a static magnetic field gradient to implement multi-qubit gates, one can also make use of microwave near-field gradients \cite{Ospelkaus2} which has been demonstrated in the pioneering work of Ospelkaus et al. \cite{Ospelkaus1}.

A challenge when using the static magnetic field gradient scheme stems from the requirement of the qubit to be made up of at least one magnetic field sensitive state. This limits the achievable coherence time and gate fidelities due to uncontrolled magnetic field fluctuations \cite{Khromova1} and measures have to be taken to shield or compensate these fluctuations. An efficient method of obtaining a qubit which is robust against magnetic field noise is by making use of microwave dressed-states \cite{Timoney1, Webster1}. These dressed-states have been shown to exhibit a coherence time three orders of magnitude longer compared to bare magnetic field sensitive qubit states and have already been combined with a static magnetic field gradient to cool a single ion to the quantum ground-state of motion using long-wavelength radiation \cite{Seb2}. Using such dressed states that give rise to quantum-engineered clock states, a high-fidelity two-qubit gate has recently been demonstrated \cite{Seb1} with the method being capable of producing fault-tolerant quantum gates.

We note that recent work has shown the possibility to cancel the carrier transition during two-qubit gate operations which is expected to permit much faster gates \cite{Timoney1, Cohen1}. For the design of a scalable quantum computer module it is highly advantageous to be able to rely on the matured and commercially developed field of microwave engineering, allowing stable microwave and RF fields to be generated at comparably low cost and, for a typical user, with a fraction of the complexity of laser systems. Furthermore, microwave radiation can naturally address a large spatial volume, making it very useful when scaling a given operation to many ions.

Static magnetic field gradient induced couplings based on the approach outlined in Ref. \cite{Seb1} will therefore be used as a basis for two-qubit gate operations within individual modules described here.

\section{Description of individual quantum computer modules}

\begin{figure}[htp]
\begin{center}
\includegraphics[width=8.6cm]{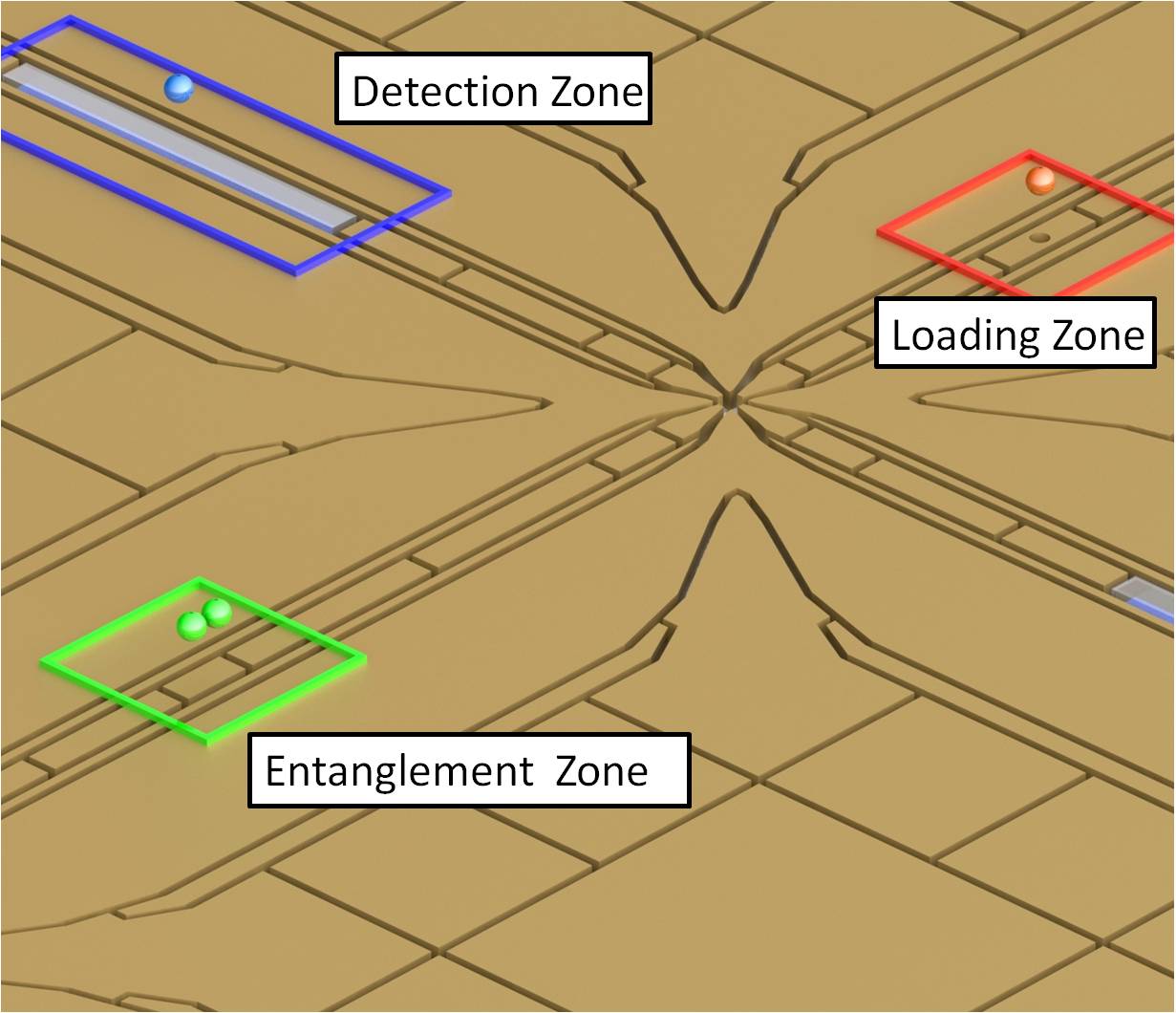}%
\caption{ X-junction featuring multiple zones, including a loading zone (marked red) in selected junctions. Multi-qubit gates are performed after bringing two or more ions (green) together in the gate zone (marked green). The gates are performed by applying a static magnetic field gradient produced by current wires placed underneath the electrodes. State readout is carried out in the readout zone (marked blue) using global laser fields and photo detectors placed underneath the electrodes.}  
\label{JunctionSequence1}
\end{center}
\end{figure}

We propose a blueprint for a scalable quantum computer module, which makes use of the discussed microwave-based multi-qubit gate scheme and is fabricated using silicon microfabrication technology. Each module is a unit cell for a large-scale quantum computer and features microfabricated ion trap X-junction arrays \cite{Stick1, Seidelin1, Hughes1}. In each X-junction two or more ions are trapped and feature up to three different zones, as shown in Fig.~\ref{JunctionSequence1}, which include a microwave-based gate zone, a state readout zone and a loading zone. Once an ion is trapped in the loading zone, high-fidelity ion shuttling operations \cite{Bowler1, Wright1} transfer the ion to the gate zone. There, ions can be individually addressed using locally adjustable magnetic fields and entangled using static magnetic field gradients in conjunction with global microwave and RF fields. When the state of the qubit needs to be detected the ion is transferred to the readout zone, where global laser fields and on-chip photo detectors are used for state readout. A second ion species is used to sympathetically cool the qubit ion without affecting its internal states \cite{Kielpinski2}. All coherent quantum operations are performed and controlled by on-chip electronics relying only on global microwave fields. In our microwave-based architecture laser light is only required for state preparation and detection and sympathetic cooling. The required laser beams do not have to be precisely aligned or individually controlled.

\begin{figure}[htp]
\begin{center}
\includegraphics[width=8cm]{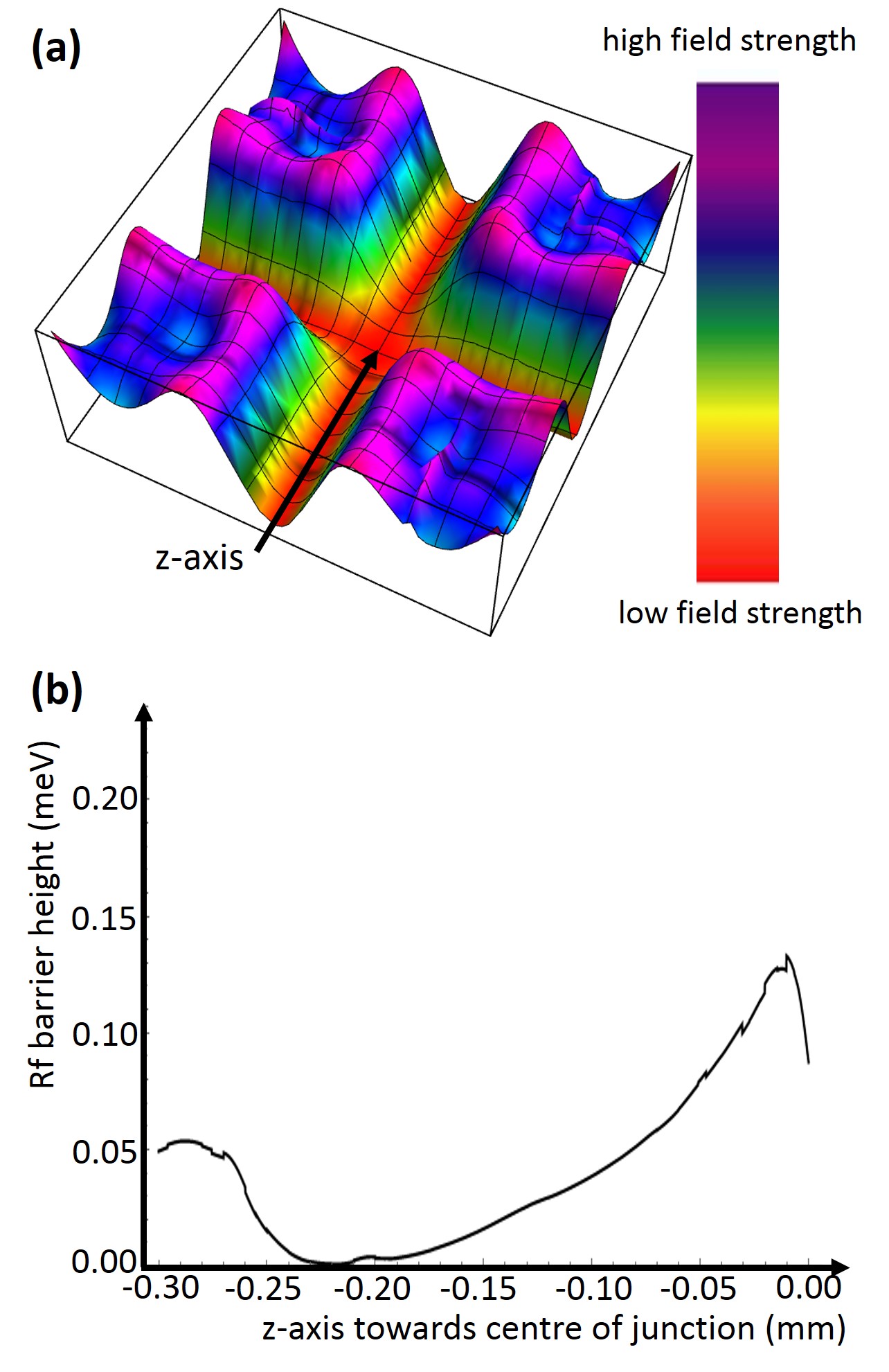}%
\caption{(a) Illustration of the RF pseudopotential at the  ion height of the proposed optimised X-junction geometry and (b) the RF barrier when moving along the RF nill in the z direction.}  
\label{JunctionSequence2}
\end{center}
\end{figure}

The design of the X-junction, individual zones, control electronic, cooling and alignment system of the modules will be described in detail in the next paragraphs. We start with the ion trap X-junction and its arms, which constitute the unit cell of the modules and are the core element of the proposed architecture. The gate, detection and loading zones, which are placed in the arms of the junction and an optimized design of the junction electrode geometry, which allows for fast high-fidelity ion shuttling and separation are essential for the operation of the modules. High-fidelity shuttling through junctions requires a highly optimized electrode geometry \cite{Hucul2, Wright1}. An example for such a junction geometry including its arms is shown in Fig.~\ref{JunctionSequence1}, featuring minimal RF barrier and barrier gradient shown in Fig.~\ref{JunctionSequence2}. The optimized electrode geometry is combined with static voltage electrodes designed for fast and efficient ion shuttling and separation \cite{Nizamani1}. The RF barrier of the highly optimized X-junction was simulated to be on the order of 0.15 meV for a trap depth of $\sim$80 meV and an ion height of 100 $\mu$m. High-fidelity shuttling through X-junctions in a surface trap with similar barrier has been successfully demonstrated \cite{Wright1}. We propose to use fast ion shuttling through the junction, similar to the work presented in \cite{Walther1, Bowler1} and fast ion separation demonstrated in \cite{Bowler1, Ruster1} to achieve a fast quantum computer cycle time.

\begin{figure}[htp]
\begin{center}
\includegraphics[width=8.6cm]{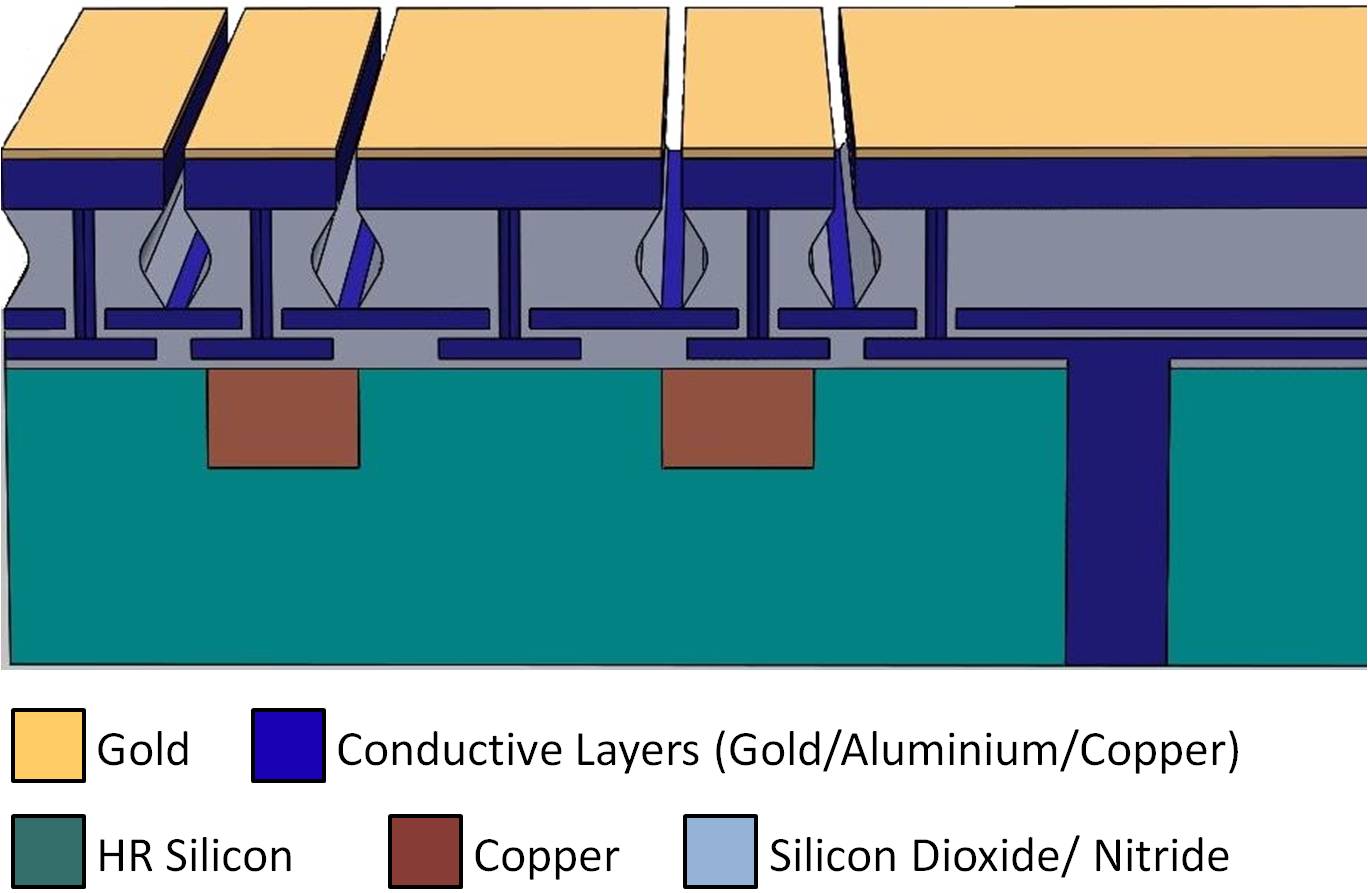}%
\caption{Layer structure of the ion trap chip consisting of highly resistive silicon substrate (HR silicon) and copper current wires embedded in the silicon. Conductive and insulating layers form buried wires, vertical interconnects (VIAs), through-silicon vias (TSVs) and electrodes.}  
\label{JunctionSequence3}
\end{center}
\end{figure}

Decoherence that would be caused by transferring ions through strong magnetic field gradients required for microwave-based quantum gates is avoided by globally turning off the gradient  fields during shuttling operations. Remaining spatially varying magnetic fields are compensated for by mapping the magnetic fields in the junctions. Slow variations of the global magnetic field can be detected using dedicated ions at various positions across the module \cite{Baumgart1} and compensated using local magnetic field coils, which will be described in more detail later. Static voltage electrodes are connected using proven and developed vertical interconnect access (VIA) and buried wire technologies \cite{Hughes1}. Additionally, a structured ground plane layer is used to avoid exposed dielectrics. The microfabricated conductive and insulating layers are placed on a high resistivity (HR) silicon substrate exhibiting minimal RF losses. The resulting layer structure is shown in Fig.~\ref{JunctionSequence3}. Using through-silicon vertical interconnect access (TSV) structures, connections to the static and RF electrodes are made from the back of the structure holding the X-junction. The electronic control systems generating the static and RF voltages will be described in more detail in the electronic control section below.

Initial loading of ions and replacement of lost ions is performed using the loading zones placed in one arm of the X-junction close to the edge of each module. Backside loading zones require a global ionization laser beam in combination with an atomic flux originating from the back of the substrate, commonly known as backside loading \cite{Amini1}. The atomic flux is generated by an atomic oven passing through slots fabricated into the silicon substrate and carefully designed centre segmented electrodes, shown in Fig.~\ref{JunctionSequence1}. When an ion is lost from a particular position within a module a new ion is trapped and all ions placed between the position of the lost ion and the loading zone are shifted by one position, requiring only single shuttling sequences.

\begin{figure}[htp]
\begin{center}
\includegraphics[width=8.6cm]{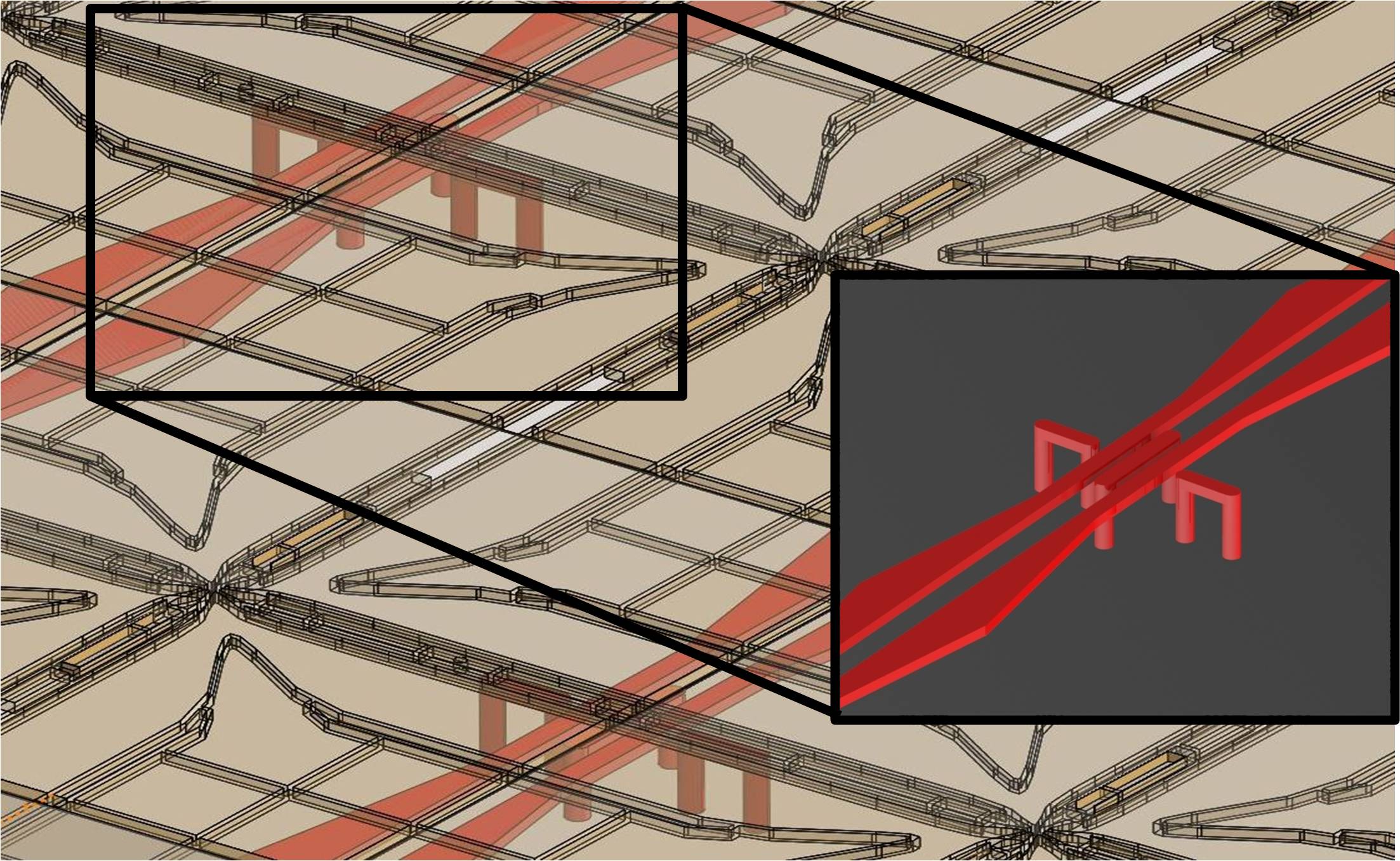}%
\caption{Illustration showing an isometric view of the two main gradient wires placed underneath each gate zone. Short wires are placed locally underneath each gate zone to form coils, which compensate for slowly varying magnetic fields and allow for individual addressing. The wire configuration in each zone can be seen in more detail in the inset.}  
\label{EntanglementZone1}
\end{center}
\end{figure}

In another arm of the optimized X-junction a gate zone is located that features a strong magnetic field gradient and an adjustable local magnetic field offset. The required magnetic field gradients and fields are generated using current carrying wires and coils embedded in the silicon substrate as shown in Fig.~\ref{EntanglementZone1}. Large static magnetic field gradients of $150\;$T/m at the ions position (100 $\mu$m above the electrode surface) are used for fast, high-fidelity microwave gates. To generate these gradients a current of $\sim$10$\;$A is passed through each copper wire. Conductivity and cooling of the silicon substrate and copper wires will be discussed in detail in the cooling system description below. The strong magnetic field gradients work in combination with global long-wavelength radiation fields to perform multi-qubit gates in parallel in the entire quantum computer architecture following the method proposed in Ref. \cite{Seb1}. Here the correlation between the number of ions and number of required gate fields vanishes. Therefore, instead of requiring thousands or even millions of individually controllable laser or microwave fields, the method used here only requires a handful of global microwave fields originating from emitters periodically placed within the vacuum chambers to implement the required quantum logic on arbitrarily many ions. This scaling method can be implemented using two different approaches. The first involves making use of the already present static magnetic field gradient in the gate zone, where the ions can be shuttled along this gradient to change the ions offset magnetic field, thereby bringing the qubit frequency into resonance with the global microwave fields of choice to perform the desired single or multi-qubit gate. The alternative approach involves using local B-field coils to bring the qubit into resonance. Since the proposed architecture already features local offset coils used to adjust the magnetic field in each gate zone and to compensate slow variations of the local magnetic field, we will focus on the latter approach to shift qubits in and out of resonance with global radiation fields.

The local local B-field coils are placed underneath each gate zone and are shown in Fig.~\ref{EntanglementZone1}. In order to ensure that they can be used to shift ions in and out of resonance with multiple global microwave fields a fast control system is implemented. The on-chip control electronics is based on digital to analogue converters (DACs) which control the currents applied to the coils with 16 bit precision and an update rate of $>$1 MHz. Required microwave frequencies for the global fields are generated by commercial frequency generators and amplifiers supply the signal with sufficient amplitude to emitters inside the system. The gate zones then perform all quantum operations controlled by in-vacuum electronics and powered by global long-wavelength radiation fields.

\begin{figure}[htp]
\begin{center}
\includegraphics[width=8.6cm]{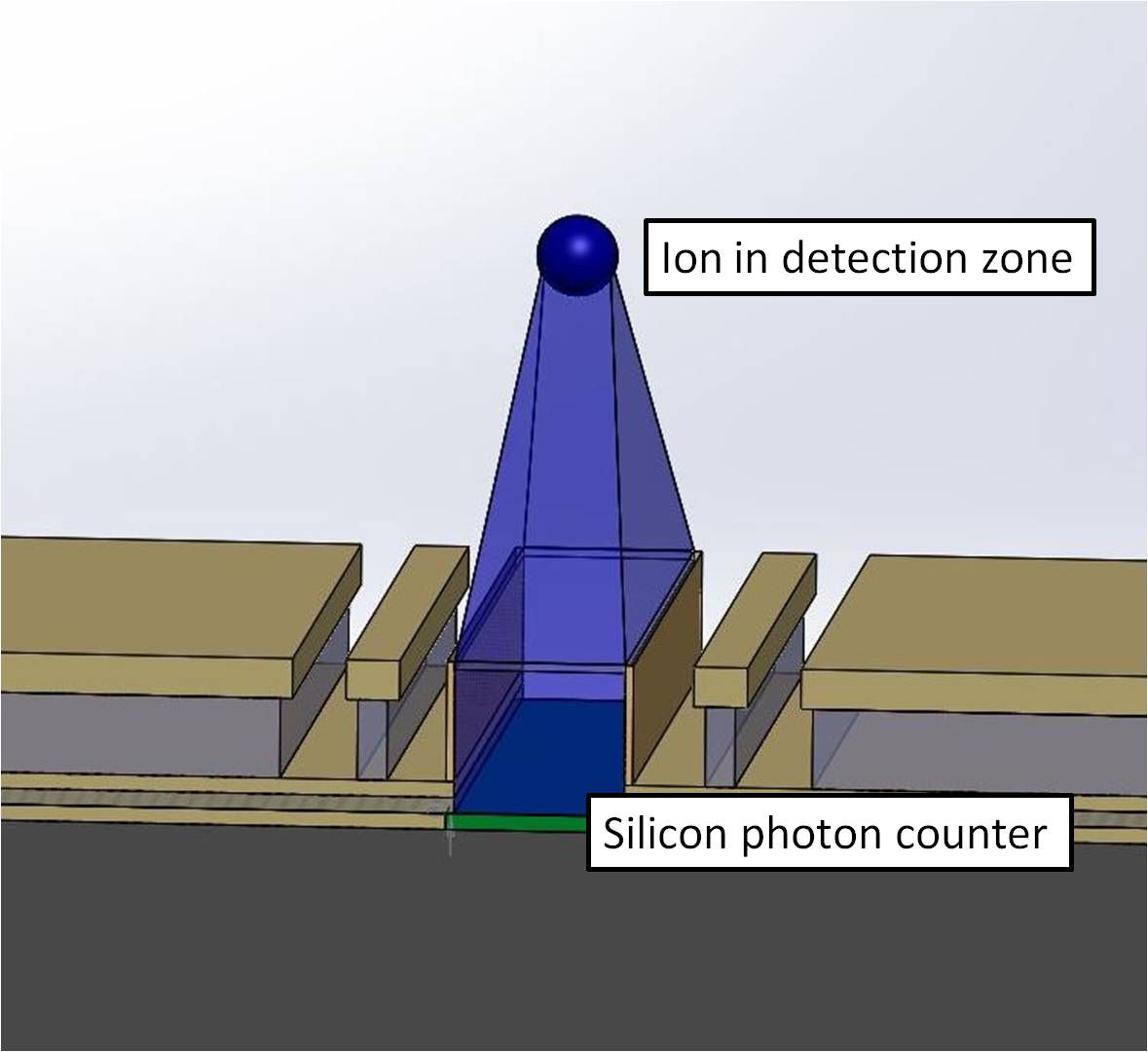}%
\caption{Silicon photo detector (marked green) embedded in the silicon substrate, transparent centre segmented electrodes and possible detection angle are shown. VIA structures are used to prevent optical crosstalk from neighbouring readout zones.}  
\label{EntanglementZone2}
\end{center}
\end{figure}

To detect the quantum state of the ions after performing single and multi-qubit gates, readout zones are incorporated into another arm of the X-junction as shown in Fig.~\ref{JunctionSequence1}. In the readout zone, multiple centre segmented electrodes are made of indium tin oxide (ITO) instead of gold. ITO is highly UV transparent and allows the light emitted from an ion placed above the zone to pass through the electrodes. Photodetectors are fabricated onto the silicon substrate and separated from the electrodes by a highly UV transparent dielectric layer, similar to the concept presented in \cite{Eltony1} and shown in Fig.~\ref{EntanglementZone2}. VIA wall structures are used to prevent optical crosstalk from neighbouring readout zones. Commercial silicon based microfabricated photon counters (Hamamatsu S12571-100, multi-pixel photon counters, MPCCs) reach quantum efficiencies of $\sim$30\% and are compatible with the proposed silicon substrate. When cooled to 77 K they also show a reduction of dark count rate on the order of $10^5$, to $\sim$1 Hz \cite{Otono1}.

The total photon detection efficiency of this detection setup will be on the order of 2\% considering an 80\% transmission rate of the ITO and dielectric layer and collection efficiency of $\sim$10\%. The detection efficiency and dark count rate are comparable to the values given in \cite{Noek1} and a similar state readout fidelity on the order of 99.9\% for a detection window of 25 $\mu$s can therefore be expected for this setup.

State readout operations will have to be performed many times during error corrected logical qubit operations. To preserve the state of physical qubits performing these operations, only ions placed inside the readout zones are illuminated while ions placed in the gate zones are not. Readout and gate zones are placed in perpendicular arms of junctions, as shown in Fig.~\ref{JunctionSequence1}. Global laser beams are steered parallel to and between the gate zone arms, which are separated by 2.5 mm, only addressing the ions in the readout zones shown in Fig.~\ref{EntanglementZone2}. The required accuracy of the beam steering is readily achieved using in-vacuum optics.

The X-junction structures equipped with the zones discussed, occupy an area of $2.5\times2.5$ mm$^2$ and can be fabricated in very large numbers on a silicon wafer to form the scalable quantum computer module. A total of 1296 individual X-junctions can be monolithically fabricated onto $90\times90$ mm$^2$ silicon wafer pieces, compatible with standard 150mm wafer sizes. If all these X-junctions are electrically connected together, the capacitance and power dissipation will become too large to be driven with a standard helical resonator of high quality factor \cite{James1}.

Simulations performed using the ADS software tool (Advanced Design System by Keysight Technologies) show that by connecting $6\times6$ junctions together to form an electrical submodule, the capacitance can be kept below 80 pF and a quality factor of Q$>200$ is achievable using a compact helical resonator of $\sim$15 mm diameter. An additional requirement to achieve a high quality factor is to use a low RF loss substrate. Therefore a highly resistive (HR) silicon substrate with a bulk resistivity of 50 k$\Omega$cm was assumed for these simulations. Compact resonators are placed inside the system underneath the module and connected with shielded cables to the electrical submodules. All resonators are attached to the same frequency source and the resonant circuits are tuned into resonance with the frequency source using variable capacitors.  Close proximity of the resonators to the electric sections and careful design of the wiring should result in a negligible phase difference between RF electrodes of different electrical sections.

Each electrical submodule features 1224 static voltage electrodes and 108 individual local gradient current wires. The required static voltages and currents are supplied by DACs inside the vacuum system similar to the concept presented in \cite{Guise1}. DACs are fabricated on separate silicon substrates, which are attached to the ion trap substrate using TSV and wafer-stacking \cite{Patti1} technology. Each wafer layer features 4 DACs with 160 analogue outputs in total (the DAC AD5370 has sufficient outputs and was used as an example, but a modified version will be required that operates at higher update rates) and combined with the required TSV and RC filters occupy an area of no more than $15\times15$ mm$^2$. Generating enough analogue outputs requires a total of 9 wafer layers that will be stacked together. An additional layer is used to house an electronic control unit, which controls the in-vacuum DACs and detection system.

Each scalable quantum computer modules is made up of $6\times6$ electrical submodules, fabricated onto a $90\times90$ mm$^2$ HR silicon wafer piece. The module is controlled by on-chip electronics and performs the required quantum operations using magnetic field gradients, local magnetic fields and global laser and microwave fields. Embedded copper wires generating the magnetic field gradients, shown in Fig.~\ref{EntanglementZone1}, are routed in such a way that only 4 high current connections are required per module.

Passing large currents of 10 A through wires with a small cross-section ($\sim30\times60$ $\mu$m$^2$) makes it essential that the resultant heat is efficiently distributed and transported away from the modules. In addition, the power dissipated by the ion trap structure and the in-vacuum electronics needs to also be transported away from the modules. Melting of the wire structures can be avoided by cooling the silicon substrates to below 100 K, which results in an extremely high thermal conductivity ($\kappa>$1000~W/(mm$^2$$\cdot$K) of silicon \cite{Glassbrenner1} and an increase of the copper conductivity by a factor of 10 \cite{Matula1}. To estimate the temperature of the copper wires in this design the total heat output per module has to be calculated. Considering the heat generated by the copper wires, RF dissipation in the trap structure and power dissipation of the on-chip electronics, a maximum heat output per module is estimated to be on the order of 1000 W, which is equal to 0.12 W/mm$^2$ and less than a modern computer processer unit (Intel Ivy Bridge 4C has a power dissipation of $\sim$0.5 W/mm$^2$). A detailed calculation will be given in the "Supplementary Materials" section. The assumed heat output is likely to be significantly lower as the power dissipation of on-chip electronics and ion trap structures are estimated for room temperature. To efficiently remove the heat from the modules a liquid nitrogen microchannel cooler is integrated into the back wafer of the modules. Deep trenches are etched into the backside of the last wafer forming channels through which liquid nitrogen is passed. The channels are covered using an additional silicon wafer. Fabricating the entire module including the liquid cooler out of silicon prevents additional stress and wafer-bow arising from different thermal expansion coefficients.

A similar microchannel cooler has been shown to achieve a heat transfer coefficient of $>$0.1~W/(mm$^2$$\cdot$K) \cite{Riddle1}, which is sufficient for this system. Based on the total heat dissipation, thermal conductivity of silicon wafers and copper interconnects, the thermal gradient between the copper wires and the coolant through the multi-wafer package and microchannel cooler will be only $\sim$2 K, preventing the copper wires from melting. Liquid nitrogen, which is cooled from 77 K to 65 K to prevent boiling inside the microchannel cooler, will be used as coolant and is supplied to the modules using multiple UHV compatible flexible steel tubes. Continuous flow liquid nitrogen coolers are commonly used for detectors and if designed correctly introduce minimal vibrations to the system \cite{Lizon1}.

\begin{figure}[htp]
\begin{center}
\includegraphics[width=8.6cm]{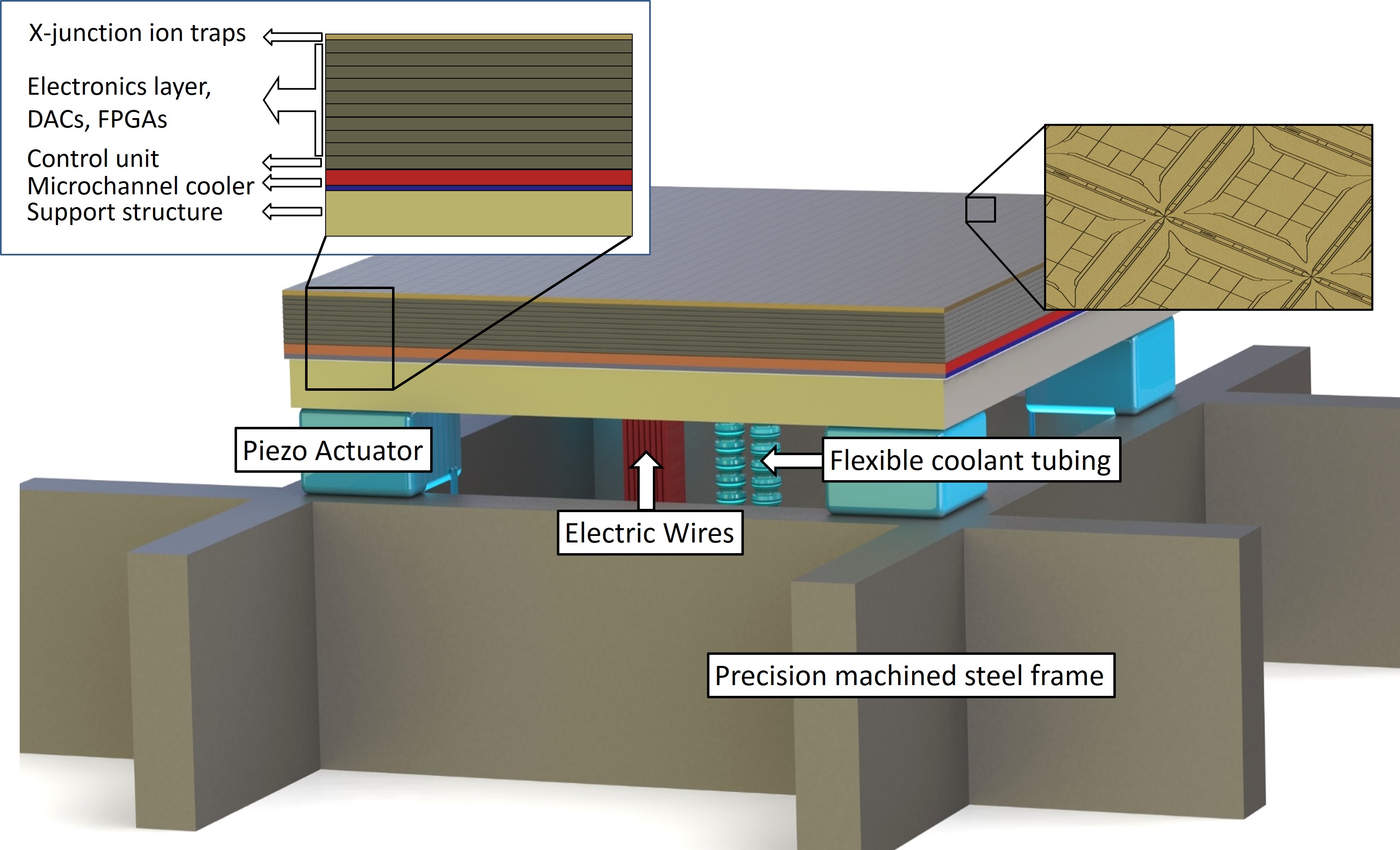}%
\caption{One module consisting of 36 x 36 junctions placed on the supporting steel frame structure: Nine wafers containing the required DACs and control electronics are placed between the wafer holding 36 x 36 junctions and the microchannel cooler (red layer) providing the cooling. X-Y-Z piezo actuators are placed in the four corners on top of the steel frame allowing for accurate alignment of the module. Flexible electric wires supply voltages, currents and control signals to the DACs and control electronics. Coolant is supplied to the microchannel cooler layer via two flexible steel tubes placed in the centre of the modules.}
\label{60squnit}
\end{center}
\end{figure}

Each of these modules can work as a stand-alone small-scale quantum processer module featuring 1296 X-junctions. If one wants to perform a computationally hard problem, such as Shor factorizing a 2048 bit number a much larger architecture, consisting of many modules will be required. Each module will have to be interfaced with each other to create a universal quantum computer architecture. The approach presented by Monroe et al. \cite{Monroe2} makes use of photonic interconnects and commercial fibres to interface arbitrarily many modules. Fibre switches can then be used to connect any module with any other module in the architecture. This approach has great potential, but the performance of such a system is currently limited by the interaction rate between modules \cite{Hucul1}, which is typically much slower than other quantum operations \cite{Noek1, Ballance1, Harty1} performed by the modules.

\section{Scaling modules to a universal quantum computer architecture}

We propose an alternative scheme that does not rely on photonic interconnects and is therefore not limited by the interaction rate of these. In our approach, modules are designed in such a way that ions can be directly shuttled from one module to another. RF and static voltage electrodes thus need to be fabricated all the way to the edge of the modules so that the electric fields confining the ions reach beyond the edges. The modules must also be accurately aligned so that two neighbouring modules create an overlapping electric field. If such an electric field can be created, ions can be shuttled from one module to another.

The resulting two-dimensional module array will feature fast interaction rates between nearest-neighbour modules without the need of a special photonic interconnect system. The challenging part of this scheme is to accurately align all modules to each other to prevent large barriers or interruptions of the overlapping electric fields to occur.

We have performed boundary element method (BEM) electric field simulations of three dimensional trap structures to investigate the feasibility of shuttling ions from one module to another taking into account the possible misalignments between adjacent modules. We have analysed the electric potential and RF barrier caused by RF rails misaligned in different directions and magnitude. Results of the simulations show that an RF barrier occurs, similar to that found in the centre of an X-junction.

\begin{figure}[htp!]
\begin{center}
\includegraphics[width=8.6cm]{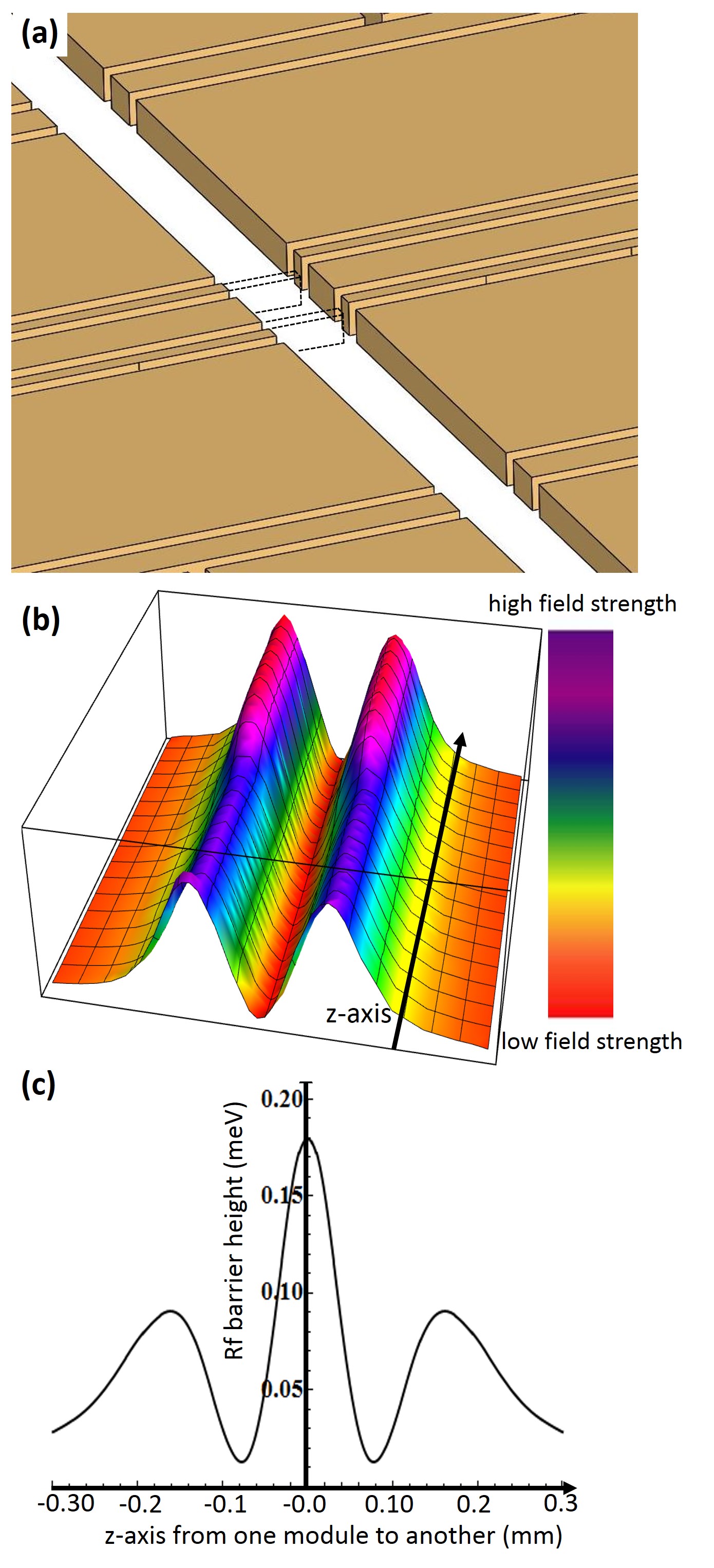}%
\caption{(a) Illustration of two modules misaligned in the x-y-z axes by 10 $\mu$m each and the corresponding RF pseudopotential (b) at the ion height. The resulting RF barrier (c) when moving along the RF nill in the z direction is given in meV. }
\label{Misaligned}
\end{center}
\end{figure}

In the case of a misalignment in all three axes by $\leq$10 $\mu$m, the simulated RF barrier was found to be $\sim$0.2 meV, as shown in Fig.~\ref{Misaligned}, for a trap depth of $\sim$100 meV and an ion height of 100 $\mu$m. The barrier of 0.2 meV is of similar height to the one found in our optimized X-junction centre and the one presented in Ref. \cite{Wright1}, where high-fidelity shuttling was successfully demonstrated. Shuttling fidelities comparable or higher than through the X-junction centres can therefore be expected if the RF voltages applied to both modules show no significant phase or frequency difference and the neighbouring modules can be aligned with $\leq$10 $\mu$m accuracy.

As discussed in the previous section, all resonators providing the RF voltages to the modules are connected to the same frequency source. In addition all modules feature the same path length and impedance of the coaxial connections from the resonators to the RF electrodes. This will result in a negligible RF phase difference of the electric fields generated by adjacent modules, which could otherwise weaken the trap depth at the intersection between modules.

To achieve an alignment of the modules with $\leq$10 $\mu$m accuracy in three dimensions, precision machined steel frames are mounted inside the vacuum chambers. The planarized top surface of the steel frames is characterised using an interferometric measurement system and modules are mounted on the surface using a high precision die bonder tool. To increase the alignment accuracy and to allow for drift compensation multiple UHV and cryogenic compatible X-Y-Z piezo actuators are placed between the bottom of each module and the top of the steel frames. The illustration in Fig.~\ref{60squnit} shows a pictorial representation of a single module of the scalable architecture including required connections and attached piezos on top of a steel frame.

The exact positions of the modules are determined using microfabricated diffraction gratings on the front of the modules in combination with a laser measurement system. The alignment system determines the position of neighbouring modules and corrects for misalignment using the piezo actuators and if required the die bonder tool. The requirements for the high-precision position measurement system and the module placement are three orders of magnitude less demanding than what is currently achieved by a lithography stepper systems, where 3-5 nm alignment precision of large wafer stages in vacuum is routinely achieved (ASML TWINSCAN NXE:3300B).

The discussed alignment capability would be severely hindered if the modules are strongly warped in random directions or the edges of the modules are not precisely fabricated. Due to the large thickness of the silicon wafer modules (on the order of 10 mm) negligible wafer bow is expected and the bow will also be characterized for all modules before assembly. The edges of the modules are created using-high resolution photolithography and anisotropic dry etching (Bosch process). These process steps, which are commonly used in the microfabrication of microchips and MEMS devices, reach sub $\mu$m precision and will not limit the alignment capabilities. In addition only the top wafer carrying the ion trap structures will feature the full footprint size of the module, all other wafers, containing control and detection electronics and cooler will have a slightly smaller footprint.

\begin{figure}[htp!]
\begin{center}
\includegraphics[width=8.6cm]{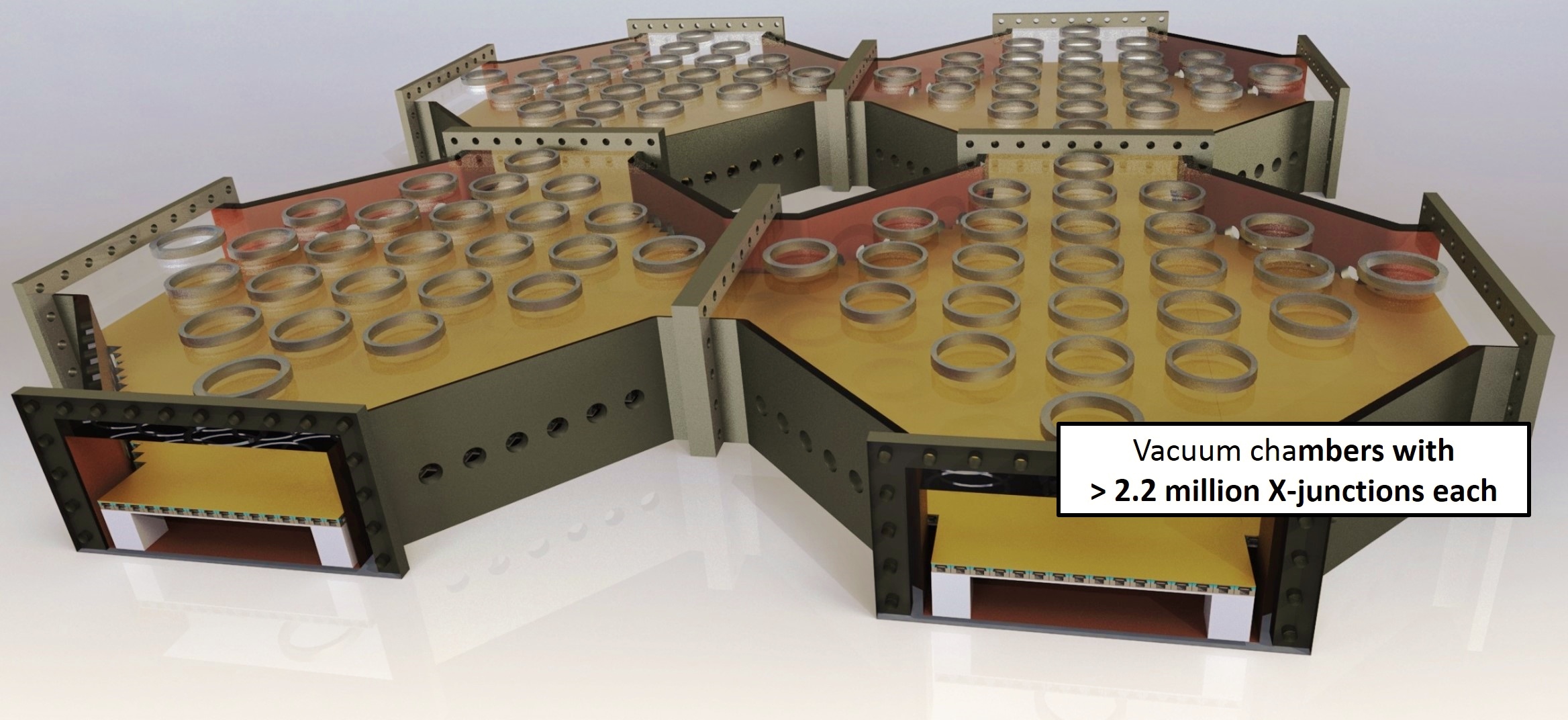}%
\caption{Schematic of octagonal UHV chambers connected together, each chamber is 4.5x4.5 m$^2$ large and can hold $>$ 2.2 million individual X-junctions placed on steel frames.}  
\label{ScalableVacuumSystem}
\end{center}
\end{figure}

The discussed steel frames are accurately placed inside large octagonal shaped UHV chambers ($\sim4.5\times4.5$ m$^2$ large), which feature viewports on the sides and top to allow for optical access, as shown in Fig.~\ref{ScalableVacuumSystem}. Imaging, beam shaping and guiding the laser light fields above the trap surfaces is achieved using in-vacuum optics. Each chamber also incorporates all required feedthroughs for currents, static voltages, RF and microwave signals, coolant and digital control signals. In addition, the chambers are equipped with liquid nitrogen cooled heat shields and a variety of vacuum pumps to create an ultra-high vacuum environment. Multiple chambers can be directly attached together creating a modular universal quantum computer of arbitrary size. Each chamber shown in Fig.~\ref{ScalableVacuumSystem} can hold $\geq$ 2.2 million individual junctions.

\section{Surface error correction code}

Performing computationally hard problems using a quantum computer, such as prime factoring of a 2048 bit number, requires logical operations to be performed with a much lower error rate than achievable by any potential quantum system and will prohibit a successful outcome of such a computation. Quantum error correction, which uses multiple physical qubits to create logical qubits with a much lower error rate, is thus a necessity for scalable quantum computing. Following the pioneering work by Steane \cite{Steane1} and Shor \cite{ShorECC}, a variety of different error correction strategies have been developed. In the original proposals, logical qubits were encoded in a number of physical qubits using special code word states that allow identification and fixing of errors (occurring on single physical qubits) without destroying the logical qubit states. These codes require the error probabilities associated with each operation on the physical qubits to be below a very challenging threshold for the error code to work and even lower error rates (of order $10^{-5}$, \cite{Steane1}) for a practical implementation with manageable resource requirements.

Since then, schemes have been developed that can tolerate much larger error probabilities, but rely on the coding of logical qubits in a larger number of physical qubits compared to the previously discussed error correction codes. One such scheme is the surface error correction code described by Fowler et al. \cite{Fowler3}, which tolerates gate error probabilities of up to $10^{-2}$ and relies only on nearest neighbour interactions. We will briefly summarize how the implementation of the surface code protects from errors, discuss how it can be implemented with this two-dimensional array architecture and how it can be used to perform fault-tolerant logic operations.

\begin{figure}[htp!]
\begin{center}
\includegraphics[width=8.6cm]{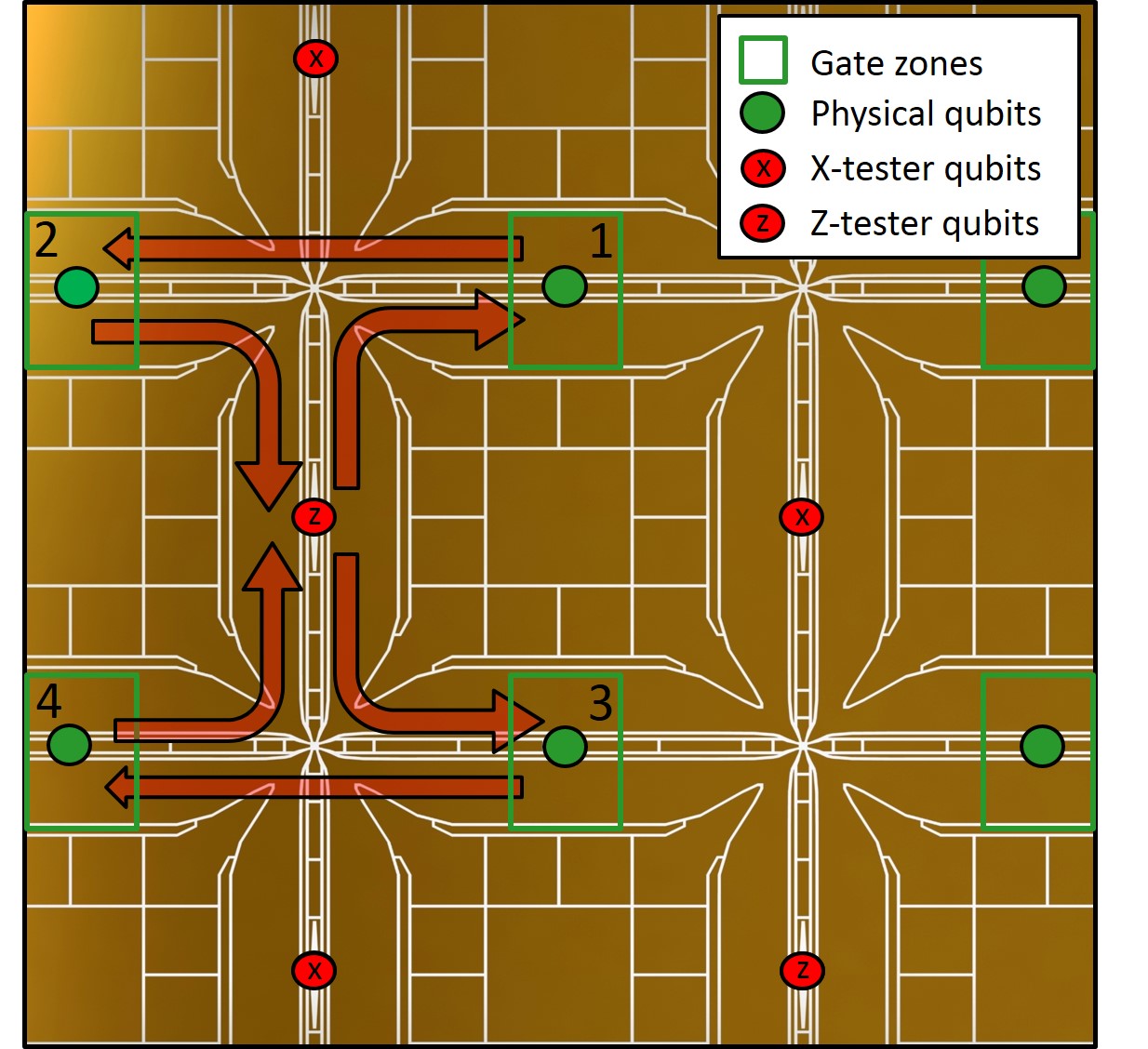}%
\caption{Small section of the scalable architecture illustrating how data and measurement qubits interact with each other in the gate zones to execute the surface code. Data qubits are static and measurement qubits are shuttled to all adjacent gate zones.}  
\label{SurfaceErrorPic}
\end{center}
\end{figure}

The surface code requires physical qubits to be placed in a regular lattice, that we can decompose into one sublattice holding so called data qubits and another sublattice holding measure-X and measure-Z qubits. In our architecture, two ions are trapped in each X-junction section, as shown in Fig.~\ref{SurfaceErrorPic}. One ion is permanently placed in the gate zone constituting the data qubit, the second ion is alternatively a measure-X or measure-Z qubit, placed in the readout zone and can be shuttled to the four adjacent data qubits.

Measure-X and measure-Z qubits constantly monitor the states of their four nearest neighbour data qubits. The measure-Z qubits perform four successive CNOT gates with the data qubits in their respective gate zones, in the order displayed in Fig.~\ref{SurfaceErrorPic}, after which the state of the measure-Z qubit is detected. The measure-X qubits perform almost the same sequence, but an additional Hadamard gate is applied to them before and after the four CNOT gates. The measure qubit sequence is run simultaneously in a synchronized manner with all measure qubits of the entire architecture and repeats itself over and over again throughout the calculation. These sequences of four CNOT gates (two additional Hadamard gates for Z-errors) are designed to perform parity checks of the fourfold $\sigma_x$ and $\sigma_z$ operators that allows for error detection on the surface code. The eigenstates of the products are dependent on the four neighbouring data qubits and all products commute with one another. If a single data qubit of the array undergoes an error it will result in eigenvalue flips of the two adjacent operators and the error is identified by the surface code at the end of the error chain. The error could then be corrected, but an easier and more robust way is to merely store the error information. The correction can then be performed `in software', i.e., by translating the measurement results at the end of execution into the appropriately modified values.

Should ions be lost from the qubit space due to entering a dark state, optical pumping is used to drive them back to the qubit space and the net effect is then equivalent to a qubit error of the form detected and corrected by the code. Complete loss of a measure qubit due to collisions with background gas is detected due to the periodic state readout of these after each measurement cycle. When such a loss is detected, the empty measure qubit site is replenished with a new ion from one of the loading zones by shifting all ions between the loading zones and the empty measure site. The loss of a data qubit cannot be detected directly, we therefore propose to use the measure qubit as an indicator if a data qubit is present in the gate zone or not.

When the measure qubit is shuttled into the gate zone, it will experience different electric potentials if a data qubit is present. If the data qubit is present, the two ions occupy left and right potential minima, while if the data qubit ion is missing, the measure qubit will occupy the centre of the potential well. When shuttling the measure qubit out of the gate zone a special potential well is applied that leaves measure qubits placed in the centre of the well in the zone, while off-center measure qubits are extracted and the measurement cycle proceeds. The measure qubit replaces missing data qubits automatically and is replenished in a later cycle using the previously discussed method. Typically after a single execution of the measurement cycle, the surface code cycle is re-established.

In an arbitrarily sized two-dimensional array of qubits protected by the surface code only a single unique quantum state exists. To implement logical qubits in such a protected array additional degrees of freedom need to be added by turning off the readout carried out by two measure-X or -Z qubits at a distance of a few sites \cite{Fowler3}.

Logical one- and two-qubit gates are implemented by so-called braiding operations, which are described in more detail in Ref. \cite{Fowler3}. Braiding operations can be implemented in our surface code protected architecture by 'switching' on and off multiple measure qubits in such a way that they form paths (braids) that lead from the measure qubits, defining one logical qubit, around the measure qubits defining another logical qubit, and back. After a complete cycle where the measure qubits are switched on again along the same path, a logical two-qubit gate is effectively implemented on the logical qubits. While performing such a logical qubit gate, the physical qubits in the array do not have to perform any additional operations then the ones required for the error correction cycle. Regions of the lattice forming the defects only have to be switched off, which corresponds to data qubits remaining in their gate zone and measure qubits are `parked' in their readout zone. The defect regions are not constant in space/time for the entire logic operations and therefore the regions will be switched on and off at different times, depending on the topological geometry of the circuit. When a region is switched back on the error correction cycle is resumed and the physical qubits perform their designated operations again.

Non-Clifford single qubit $\pi/8$ Z-rotations are the most challenging operations to implement with the surface code. In Ref. \cite{Fowler3}, it is proposed to implement these single-qubits gates by performing logical two-qubit CNOT gates using ancilla qubits which are logical qubits initialized in states of the form $|0\rangle_L + e^{i\pi/4}|1\rangle_L$. Logical qubits can not be directly initialized to such a state performing only logical gate operations. It was therefore proposed \cite{Fowler3} to perform the required rotations on the physical state of one data qubit. The data qubit state is then `injected' into an error corrected logical qubit. Simplistically, a logical qubit consisting of only one data qubit is created and the rotation is performed on the data qubit state defining the logical qubit state. Afterwards the logical qubit is `grown' to achieve the desired fault tolerance again. Although the initialized state is now error protected, the original operation creating the state was performed on a physical qubit and does not have a low enough error probability for performing computationally hard problems. Therefore it is necessary to create multiple logical qubits with the same injected states and to use an error correction code like the Steane code to distill them to high-fidelity. The Steane code is implemented using the surface code logical qubit operations and produces logical ancilla qubits, which are needed in large numbers. While their production with adequate precision is the most challenging operation (it is estimated to occupy more than 90\% of all physical qubits of the system \cite{Fowler3}), the physical qubits do not have to perform any additional or more complex operations compared to the standard error correction cycle.

By implementing and amending the surface code scheme as outlined above to accommodate one measure and one data qubit ion in each X-junction of our architecture, we have the necessary ingredients to employ the scheme analyzed in detail in \cite{Fowler3}. We assume the same gate and memory error probability of 0.1\%, which can be achieved when implementing the microwave gate scheme \cite{Seb1} with the proposed magnetic field gradient. Based on the same scheme we can give quantitative estimates on the system size and processing time for a machine that solves a relevant, hard problem, such as the Shor factoring of a 2048 bit number. For the calculations we assume a single qubit gate time of 2.5~$\mu$s, a measurement time of 25~$\mu$s and total error correction cycle time, which includes shuttling and two qubit gates times, of around 125$\mu$s. Based on these numbers, performing a 2048 bit number Shor factorisation will take on the order of 110 days and require a system size of $10^{9}$ trapped ions. Shor factoring of a 1024 bit number will take on the order of 14 days. Both of these factorizations will require almost the same amount of physical qubits as the required pace of the ancilla qubit generation is the same for a 2048 bit and a 1024 bit factorization. Trapping $10^{9}$ ions will require $15\times15$ vacuum chambers occupying an area of ca $67.5\times67.5\;$m$^2$.

We believe that these numbers are very encouraging, and we are confident that further improvements to the error correction code could bring down the overhead of performing such a calculation by up to an order of magnitude. Assuming that it will also be possible to reduce the error rate of each quantum operation below 0.01\%, it would be possible to perform the 2048 bit number factorization in approximately 10 days, requiring on the order of $3\cdot10^{8}$ ions. In addition one could implement  medium-range shuttling, of approximately 30 junctions, which could lead to a further reduction of the required qubits to $3\cdot10^{6}$ ions or $1.5\cdot10^{6}$ X-junctions. These improvements could be implemented without major changes to the modules or the architecture and all required qubits would fit into four vacuum chambers.

\section{Conclusion}

The presented blueprint for a scalable trapped-ion quantum computer module combines the advantages of microwave-based quantum gates with on-chip control electronics, which not only generate voltages to perform shuttling operations but also control all quantum information operations. When placed in a vacuum system, which supplies the modules with static and RF voltages, coolant and global laser and long-wavelength radiation fields, each module can work as a small stand-alone quantum computer. Local adjustable magnetic fields in combination with a small number of global long-wavelength radiation fields can be used for selective addressing of arbitrarily many qubits in parallel \cite{Seb1}. This has the critical advantage of replacing thousands or millions of laser beams required in previously proposed architectures \cite{Monroe2, Monroe3} with only a few global microwave fields, making this technology an efficient engineering solution to construct a large-scale universal quantum computer.

To go beyond a single module and scale to an arbitrarily large quantum computer we propose an alignment system which makes it possible to accurately align modules with their adjacent ones. The resulting system features nearest neighbour ion-ion interactions spanning across the entire architecture, making it suitable for the surface error correction code. By adding photonic interconnect regions, which poses a minor adjustment to the blueprint, the modules can also be used in alternative scalable quantum computing architectures.

We have described an engineering blueprint for a microwave-based trapped-ion quantum computer based on modules that are connected via ion shuttling, forming a scalable quantum computer architecture. The proposed modules combine the advantages of microwave-based quantum gates with local offset B-fields to remove the correlation between the number of qubits in the quantum computer and the required number of radiation fields to perform the quantum logic operations. We have shown how the surface error correction code can be implemented in this architecture and have given an estimate of the system size and execution time required to perform a 2048 and 1024 bit number Shor factorization in this scalable architecture.

\section{Acknowledgements}

We thank J. Randall,  D. Murgia and S. Webster for helpful discussions and comments. This work is supported by the U.K. Engineering and Physical Sciences Research Council  [EP/G007276/1, the UK Quantum Technology hub for Networked Quantum Information Technologies (EP/M013243/1), the UK Quantum Technology hub for Sensors and Metrology (EP/M013243/1)], the European Commission’s Seventh Framework Programme (FP7/2007-2013) under Grant Agreement No. 270843 (iQIT), the Army Research Laboratory under Cooperative Agreement No. W911NF-12-2-0072, the US Army Research Office Contract No. W911NF-14-2-0106, the Villum Foundation, and the University of Sussex. S. J. Devitt acknowledges support from the JSPS Grant-in-aid for Challenging Exploratory Research and JSPS KAKENHI Kiban B 25280034. The views and conclusions contained in this document are those of the authors and should not be interpreted as representing the official policies, either expressed or implied, of the Army Research Laboratory or the U.S. Government. The U.S. Government is authorized to reproduce and distribute reprints for Government purposes notwithstanding any copyright notation herein.

\bibliography{Scalable}

\section{Supplementary Materials}
The electric potential experienced by the ion when shuttled through junctions and between physically separate modules (Fig.~\ref{Misaligned}) was simulated using the `Essential Numerical Tools Packages` by Kilian Singer and using Wolfram Mathematica. Calculations for system sizes and calculation speeds using the error correction code were based on the work presented in \cite{Fowler3}.

Calculations for power dissipation and thermal gradients of the scalable modules were performed, assuming a 10 A current is passed through the narrowest wire section $30\times60$ $\mu$m$^2$. The wires are made of copper, for which the electrical resistivity is assumed to be 2 n$\Omega \cdot$m \cite{Matula1} and the thermal conductivity to be $\kappa \sim$482~W/(m$\cdot$ K) at 100 K. The copper wires are embedded in a silicon substrate with a thermal conductivity of $\kappa\sim$1000$\;$W/(m$\cdot$ K) at 100 K \cite{Glassbrenner1} using a thin (50 nm) titanium layer for adhesion with a thermal conductivity of $\kappa\sim$ 31$\;$W/(m$\cdot$ K) at 100 K. Assuming the heat spreads inside the 10 mm thick silicon stacked wafers at an angle of at least 36.5  degree \cite{Vermeersch} the temperature difference between the copper wires and the micro-channel made of silicon placed at the bottom of the stacked wafers was calculated to be less than 1 K.

Before one can calculate the total temperature difference between the copper wires and coolant we need to determine the total power dissipation that has to be cooled by the microchannel cooler. To calculate the power dissipated by the currents passing through one X-junction we assumed that there are two copper wires with an average width of 125 $\mu$m,  a height of 30 $\mu$m and a length of 2.5 mm. The resulting total power dissipation contribution from wires at 100 K is $\sim$ 0.3 W for one X-junction section and $\sim$ 350 W per module. The power dissipation of the DACs and electronics were estimated based on room temperature values given for an example DAC, AD5370 ($\sim$ 0.3 W) and result in an additional $\sim$ 350 W of power dissipation. Control electronics and RF power dissipation of each module was estimated to be $\sim$ 300 W in total, based on power dissipation values for FPGAs and RF simulations. The total power dissipation will therefore be on the order of 1000 W and the power density per module will be $\sim$ 0.12 W/mm$^2$, which is significantly lower then a modern CPU (Intel Ivy Bridge 4C has a power dissipation of $\sim$0.5 W/mm$^2$). All of these estimates are based on room temperature values and are expected to be significantly lower when operated at $<$100~K.

Assuming the built-in micro-channel cooler achieves a comparable heat transfer coefficient between cooler and coolant to the one published in Ref. \cite{Riddle1} of $>$0.1~W/(mm$^2$$\cdot$K) the total temperature difference will be lower than 2 K. Based on the anticipated temperature of the coolant of $<$70~K, the resulting surface temperature of the module is expected to be $\sim$72~K.

\end{document}